\documentclass[runningheads,a4paper]{llncs}

\usepackage{amssymb}
\usepackage{graphicx}

\usepackage{url}
\urldef{\mailsa}\path|{sroy7}@buffalo.edu|

\usepackage{hyperref}
 \hypersetup{colorlinks=false}
\usepackage{graphicx}
\newcommand{\ignore}[1]{}

\newcommand{\NP}{\mathrm{NP}}
\renewcommand{\epsilon}{\varepsilon}
\usepackage{amssymb}
\usepackage{graphicx}
\usepackage{verbatim} 
\usepackage{url}
\usepackage[lined,algonl,boxed]{algorithm2e}
\usepackage{hyperref}
 \hypersetup{colorlinks=false}
\usepackage{graphicx}
\begin{document}

\mainmatter  % start of an individual contribution

\title{An FPTAS for the Lead-Based Multiple Video Transmission (\textsc{LMVT}) Problem}

\titlerunning{An FPTAS for \textsc{LMVT Problem}}

\author{
Swapnoneel Roy \and Atri Rudra%
\thanks{Research supported in part by NSF grant CCF-0844796.}%
}
\authorrunning{Roy S and Rudra A}

\institute{Department of Computer Science and Engineering,\\
University at Buffalo, The State University of New York,\\
Buffalo, New York 14260.\\
$\{$atri, sroy7$\}$@buffalo.edu}


\toctitle{An FPTAS for the Lead-Based Multiple Video Transmission (\textsc{LMVT}) Problem}
\tocauthor{Contents}
\maketitle
% or an extended abstract of the paper has appeared elsewehere}
%%%%%%%%%%%%%%%%%%%%%%%%%%%%%%%%%%%%%%%%%%%%%%%%%%%%%%%%%%%%%%%%%%%%%%%%%%%

%% the abstract has to PRECEDE the command \maketitle:
%% be sure not to issue the \maketitle command twice!

\begin{abstract}
  \noindent The Lead-Based Multiple Video Transmission (\textsc{LMVT})  problem is motivated by applications in managing the quality of experience (QoE) of video streaming for mobile clients. In an earlier work, the \textsc{LMVT} problem has been shown to be $\NP$-hard for a specific bit-to-lead conversion function $\phi$. In this work, we show the problem to be $\NP$-hard even if the function $\phi$ is linear. We then design a fully polynomial time approximation scheme (FPTAS) for the problem. This problem is exactly equivalent to the {\sc Santa Clause Problem} on which there has been a lot of work done off-late.
\end{abstract}

%\maketitle

%% start the paper here:
\section{Introduction}
\noindent
The \textsc{LMVT} problem deals with multiplexing videos simultaneously slot by slot over a wireless channel. Each slot could be allocated to at most one video. The number of bits that can be transmitted to a particular video in a given slot is known. Hence the videos receive variable number of bits over the various slots. The time of transmission is divided over a number of {\em epochs}. Each such epoch has a fixed number of slots (say $B$) known beforehand. The goal is to allocate all the slots in an epoch to the $n$ videos in a manner which maximizes the minimum number of bits received by any video in that epoch.

\noindent
The {\em lead} of a video is defined as the amount of time it can be played without interruption. An interruption in the playing of a video occurs when there are no more frames left  in the buffer to be played. The lead of any video is calculated at the end of an epoch using a function $\phi$ on the number of bits $b$ received by that video in that epoch. The motivation behind studying this problem is to develop a slot allocation algorithm to ensure the uninterrupted play of each video in the network. This problem has been well studied, and a number of papers have been published on it recently~\cite{paper02}, ~\cite{paper03}, ~\cite{paper04}, ~\cite{paper05}.
%\newpage
\section{Preliminaries}
In this section we present a theoretical formulation of the problem. For video $v_i$ and slot $j$ we define the {\em bit rate} $r_{ij}$ to be the maximum number of bits that can be transmitted to $v_i$ in $j$. The {\em decision} version of the \textsc{LMVT} problem can be presented as follows.

\noindent 
%\begin{algorithm}[H]
%\SetLine
%\begin{center}
%{ \textsc{LMVT} \textsc{Problem}}
%\end{center}
%\textsc{Input:} $n$ videos in the channel, $B$ slots to in the epoch, the bit rates $r_{ij} \in \mathbb{Z}^{+}$ for the $n$ videos and $B$ slots, a function $\phi(b)$ for calculating the lead for $b$ bits, a number $k$.\\
%\textsc{Question:} Is there an allocation of the $B$ slots to the $n$ videos so each video has a lead of at least $k$?
%\label{LMVT}
%\end{algorithm}

\begin{center}
\framebox{
\begin{tabular}{c}
\noindent\textsc{LMVT Problem}
\\\noindent {\sc Input:} $n$ videos in the channel, $B$ slots in the epoch,
\\ the bit rates $r_{ij} \in \mathbb{Z}^{+}$ for the $n$ videos and $B$ slots, 
\\a function $\phi(b)$ for calculating the lead for $b$ bits, and $k \in \mathbb{Z}^{+}$.
\\ \noindent {\sc Question:} Is there an allocation of the $B$ slots to the
\\ $n$ videos so each video has a lead of at least $k$?
\end{tabular}
}
\end{center}

\noindent
In~\cite{paper01}, the \textsc{LMVT} problem has been shown to be $\NP$-Hard for a specific function $\phi$ to calculate the lead based on the number of bits received. A natural greedy algorithm has been designed and has been shown to perform well in practice with experimental results. In this work, we show the problem to remain $\NP$-Hard even for the case in which $\phi$ is linear. Next we design an FPTAS for the problem.  

\section{\textsc{LMVT} problem remains $\NP$-Hard even for a linear $\phi$}
We assume a linear function $\phi$ to calculate the lead based on the number of received $b$. Let us consider a monotonic linear $\phi$, such that $\phi(b) = b$. Then the problem becomes of finding a slot allocation to maximize the minimum number of bits received by any video. We show the problem to remain $NP$-Hard even then. 

\noindent
We reduce the \textsc{Partition} \textsc{Problem}, a known $\NP$-Hard problem to \textsc{LMVT}. The \textsc{Partition} \textsc{Problem} can be presented as:

\noindent 
%\begin{algorithm}[H]
%\SetLine
%\begin{center}
%{ \textsc{Partition} \textsc{Problem}}
%\end{center}
%\textsc{Input:} A set $S \subseteq \mathbb{Z}^{+}$ with $\sum_{x \in S}x = U$.\\
%\textsc{Question:} Can $S$ be partitioned to $S^{\prime}$ and $S \setminus S^{\prime}$ such that $\sum_{x \in S^{\prime}}x = \sum_{x \in S \setminus S^{\prime}}x = U/2$?
%\label{LMVT}
%\end{algorithm}

\begin{center}
\framebox{
\begin{tabular}{c}
\noindent {\sc Partition Problem}
\\\noindent {\sc Input:} A set $S \subseteq \mathbb{Z}^{+}$ with $\sum_{x \in S}x = U$.
\\ \noindent {\sc Question:}  Can $S$ be partitioned to $S^{\prime}$ and $S \setminus S^{\prime}$
\\ such that $\sum_{x \in S^{\prime}}x = \sum_{x \in S \setminus S^{\prime}}x = U/2$?
\end{tabular}
}
\end{center}

\noindent
For the reduction, consider any instance of the \textsc{Partition} \textsc{Problem} with $S = \{x_1, x_2, \cdots, x_B\}$, $\sum_{x \in S}x = U$, and $|S| = B$. Now consider an instance of \textsc{LMVT} where we have $2$ videos $v_1$ and $v_2$, $B$ slots, and the bit rates $r_{1j} = r_{2j} = x_j \in S$, for each slot $j$. Note that $|S| = B = \# $ {\em of slots for the \textsc{LMVT} instance}. Set $k=U/2$.  

\begin{lemma}
The above instance of the \textsc{Partition} \textsc{Problem} has a solution iff the instance of \textsc{LMVT} has a solution. 
\end{lemma}

\proof
\noindent
We show that the instance of the \textsc{Partition} \textsc{Problem} has a solution iff we can find a slot allocation for the $2$ videos of the L\textsc{MVT} instance, such that the number of bits received by each video is exactly $U/2$. 

\noindent
Suppose the \textsc{Partition} \textsc{Problem} has a solution. That is we have $S^{\prime}$ and $S \setminus S^{\prime}$ such that $\sum_{x \in S^{\prime}}x = \sum_{x \in S \setminus S^{\prime}}x = U/2$. We find a slot allocation of the $B$ slots which allocates slot $i$ to video $v_1$ if $x_i \in S^{\prime}$. Else $i$ is allocated to $v_2$. We note that all the $B$ slots get allocated this way, since $|S| = B$. Now since $\sum_{x \in S^{\prime}}x = \sum_{x \in S \setminus S^{\prime}}x = U/2$, it is easy to see that the number of bits received by $v_1$ and $v_2$ is exactly equal to $U/2$.

\noindent
For the other way, suppose we have a slot allocation such that number of bits received $v_1$ and $v_2$ $=$ $U/2$. We partition $S$ in the following way: 
\noindent
\begin{enumerate}
\item If slot $i$ is allocated to $v_1$, then $x_i \in S^{\prime}$. 
\item Else $x_i \in S^{\prime} \setminus S$.
\end{enumerate}
Since $\sum_i r_{1i} = \sum_i r_{2i} = U/2$, we have $\sum_{x \in S^{\prime}}x = \sum_{x \in S \setminus S^{\prime}}x = U/2$, and hence a solution to the instance of \textsc{Partition} \textsc{Problem}.
\qed

\begin{corollary}~\cite{paper01}
The \textsc{LMVT Problem} is easy for a constant bit rate for all the $n$ videos and $B$ slots.
\end{corollary}

\proof
The \textsc{Partition Problem} has been reduced to LMVT. It is easy to see that an instance of \textsc{Partition Problem} where all the integers in $S$ are constant (equal) is easy to solve. Analogously, the instance of LMVT with constant (equal) bit rates is also easy to solve. 
\qed

\section{An FPTAS for \textsc{LMVT}}
In~\cite{paper01}, an exact dynamic programming algorithm has been designed for \textsc{LMVT}. The runtime of the exact algorithm is pseudo-polynomial in terms of the inputs. Here we describe the exact algorithm and then discretize the algorithm to design an FPTAS. 
\subsection{The Exact Dynamic Programming Algorithm}
Define ${b_i}^{max}$ to be the maximum number of bits that video $v_i$ . In other words, ${b_i}^{max} = \displaystyle\sum\limits_{j=1}^B r_{ij}$, is the number of bits $v_i$ would receive, if {\em all} the $B$ slots are allocated to it.
\noindent
Given $m$ slots $n$ videos, a $Tx$ (transmission) vector is an {\em n-tuple} $<b_1, \cdots, b_n>$ which tells us whether a slot allocation is possible such that video $v_i$ receives at least $b_i$ bits in the allocation. The length of the $Tx$ vector is the number of videos $n$. We define the predicate $F(m, T)$ for $m$ slots and $Tx$ vector $T$. $F(m, T)$ is true iff an allocation is possible to achieve $Tx$. 
\noindent
For two $Tx$ vectors $T_1$, and $T_2$, we define $T_1 \preceq T_2$ iff $T_1[i] \leq T_2[i]$, $\forall i$. It is easy to see, if $F(m, T_2)$, then $F(m, T_1)$. In the dynamic programming, we generate $\displaystyle\prod\limits_{i=1}^n ({b_i}^{max}+1)$ possible $Tx$ vectors starting from $<0, \cdots, 0>$ till $<{b_1}^{max}, \cdots, {b_n}^{max}>$. For each video $v_i$, we have the values taken from the set $\{0, 1, \cdots, {b_i}^{max}-1, {b_i}^{max}\}$.
%\newpage
\noindent
We maintain an $\displaystyle\prod\limits_{i=1}^n ({b_i}^{max}+1)$ by $n$ matrix of the vectors during the execution of the dynamic programming algorithm. Also, we have a {\em truth value} vector of length $\displaystyle\prod\limits_{i=1}^n ({b_i}^{max}+1)$. Each cell in the true value vector corresponds to the value of $F(m, T)$ for $m$ slots, and $Tx$ vector $T$. We initialize the truth value of $<0, \cdots, 0>$ to $true$ and the rest to $false$. This signifies that we can always achieve vector $<0, \cdots, 0>$, even without any slot allocation.
%\pagebreak
\noindent
We then start from $m$ $=$ $1$ to the total number of slots $B$, and evaluate the truth values. The truth values ($F(B, T)$)  at the end tell us whether that vector $T$ was achievable by a slot allocation with the $B$ slots. We then choose the vector with the maximum minimum $b_i$ value as our solution, and have the corresponding slot allocation as the optimal answer. 
\noindent
The way to evaluate $F(m, T)$ is as follows:
\begin{enumerate}
\item If $F(m-1, T)$, then $F(m, T)$.
\item Else let $W_i$ be the vector where all the positions of $W_i$ except $W_i[i]$ is equal to $T$. $W_i[i]$ $=$ $max(0, T[i]-r_{im})$. For $i$ $=$ $1$ to $n$, if $F(m-1, W_i)$, then $F(m, T)$.
\end{enumerate}
%\pagebreak

We present the whole algorithm in Algorithm~\ref{algo-dyn-prog}. 
\newpage
\begin{algorithm}[H]
%\SetLine
\KwIn{$n$, the number of videos, $B$, the number of slots, $r_{ij}$, the rate of video $v_i$ for slot $j$}
\KwOut{An allocation of the $B$ slots over $n$ videos where the minimum number of bits received by a video is maximized}
Generate the $\displaystyle\prod\limits_{i=1}^n ({b_i}^{max}+1)$ vectors where ${b_i}^{max} = \displaystyle\sum\limits_{j=1}^B r_{ij}$\;
Construct the $\displaystyle\prod\limits_{i=1}^n ({b_i}^{max}+1)$ by $n$ matrix of the vectors\;
Have the truth value vector of length $\displaystyle\prod\limits_{i=1}^n ({b_i}^{max}+1)$\;
Initialize the truth value of $<0, \cdots, 0>$ to $true$ and the rest to $false$\;

\For{$m$ = $1$ to $B$}
{
  \ForEach{$Tx$ vector $T$}
  {
    \If{$F(m-1, T)$}
    {
      Set $F(m, T)$ to $true$\;
    }%End If{x is less...}
    \ElseIf{}
    {
      \For{$i$ $=$ $1$ to $n$}
      {
        \If{$F(m-1, W_i)$}
        {
          Set $F(m, T)$ to $true$\;
        }  
      }
    }
  }
}

Return the $Tx$ vector $T$ with the maximum $min(b_i)$ value and with $F(B, T)$ true\;
\caption{The exact dynamic programming algorithm}
\label{algo-dyn-prog}
\end{algorithm}
\noindent
Algorithm~\ref{algo-dyn-prog} has a runtime of $O(Bn\displaystyle\prod\limits_{i=1}^n ({b_i}^{max}+1))$. Since $\displaystyle\prod\limits_{i=1}^n ({b_i}^{max}+1)$ can be exponentially large, Algorithm~\ref{algo-dyn-prog} has an exponential runtime. 
%\pagebreak
%\newpage
\subsection{The FPTAS}
\noindent
In the FPTAS, instead of considering all the values in the set $\{0, 1, \cdots, {b_i}^{max}-1, {b_i}^{max}\}$ for each video $v_i$, we discretize the set to the powers of $1+\varepsilon$, where $\varepsilon>0$. We define the function $\psi(i)=\lfloor(1+\varepsilon)^{\lfloor log_{1+\varepsilon}i \rfloor}\rfloor$. Now we have the set for each video $v_i$, as $\{0, \psi(1), \cdots, \psi({b_i}^{max}-1), \psi({b_i}^{max})\}$. Clearly, we have at most $log_{1+\varepsilon}({b_i}^{max}+1)$ values in the set. Hence we would have only $\displaystyle\prod\limits_{i=1}^n log_{1+\varepsilon}({b_i}^{max}+1)$ $Tx$ vectors to evaluate truth value for, in the FPTAS. In the evaluation of $F(m, T)$, instead of considering $W_i$ as in Algorithm~\ref{algo-dyn-prog}, we consider $W_i^{\prime}$ where $W_i^{\prime}$ is the vector where all the positions of $W_i^{\prime}$ except $W_i^{\prime}[i]$ is equal to $T$. $W_i^{\prime}[i]$ $=$ $max(0, \psi(T[i]-r_{im}))$. We present the FPTAS in Algorithm~\ref{fptas}.

\begin{algorithm}[H]
%\SetLine
\KwIn{$n$, the number of videos, $B$, the number of slots, $r_{ij}$, the rate of video $v_i$ for slot $j$}
\KwOut{An allocation of the $B$ slots over $n$ videos where the minimum number of bits received by a video is maximized}
Generate the $\displaystyle\prod\limits_{i=1}^n log_{1+\varepsilon}({b_i}^{max}+1)$ vectors where ${b_i}^{max} = \displaystyle\sum\limits_{j=1}^B r_{ij}$\;
Construct the $\displaystyle\prod\limits_{i=1}^n log_{1+\varepsilon}({b_i}^{max}+1)$ by $n$ matrix of the vectors\;
Have the truth value vector of length $\displaystyle\prod\limits_{i=1}^n log_{1+\varepsilon}({b_i}^{max}+1)$\;
Initialize the truth value of $<0, \cdots, 0>$ to $true$ and the rest to $false$\;

\For{$m$ = $1$ to $B$}
{
  \ForEach{$Tx$ vector $T$}
  {
    \If{$F(m-1, T)$}
    {
      Set $F(m, T)$ to $true$\;
    }%End If{x is less...}
    \ElseIf{}
    {
      \For{$i$ $=$ $1$ to $n$}
      {
        \If{$F(m-1, W_i^{\prime})$}
        {
          Set $F(m, T)$ to $true$\;
        }  
      }
    }
  }
}

Return the $Tx$ vector $T$ with the maximum $min(b_i)$ value and with $F(B, T)$ true\;
\caption{The FPTAS for \textsc{LMVT Problem}}
\label{fptas}
\end{algorithm}
\noindent
Algorithm~\ref{fptas} considers only $\displaystyle\prod\limits_{i=1}^n log_{1+\varepsilon}({b_i}^{max}+1)$ to evaluate out of the $\displaystyle\prod\limits_{i=1}^n ({b_i}^{max}+1)$ $Tx$ vectors evaluated by Algorithm~\ref{algo-dyn-prog}.

\begin{lemma}
The error generated by the rounding of $W_i$ in Algorithm~\ref{algo-dyn-prog} to $W_i^{\prime}$ in Algorithm~\ref{fptas} is at most $\frac{1}{1+\gamma}$ where 
$\gamma>0$.
\end{lemma}
\proof Suppose we have $<b_1, \cdots, b_n>$ $\Longleftrightarrow$ $<c_1, \cdots, c_n>$ from the $F(m, T)$ evaluation step of Algorithm~\ref{algo-dyn-prog}. In other words $F(m, T_1)$ for $T_1$ $=$ $<c_1, \cdots, c_n>$ has been evaluated to $true$ because $F(m-1, T_2)$ had been evaluated to be $true$ for $T_2$ $=$
$<b_1, \cdots, b_n>$. Hence, $\exists i \in [n]$, such that, $T_2[i] = b_i = max(0, c_i - r_{im})$, and for all other positions, $j$ we have $T_2[j] = T_1[j]$.

\noindent
Now suppose we have $T^{\prime}_2$ $=$ $<b^{\prime}_1, \cdots, b^{\prime}_n>$ and $T^{\prime}_1$ $=$ $<c^{\prime}_1, \cdots, c^{\prime}_n>$ in the table for Algorithm~\ref{fptas}, where $b^{\prime}_i$ $=$ $\psi(b_i)$, and $c^{\prime}_i$ $=$ $\psi(c_i)$, $\forall i \in [n]$. We want to show that if  Algorithm~\ref{fptas} evaluates $F(m, T^{\prime}_1)$ to $true$ if $F(m-1, T^{\prime}_2)$ was evaluated to $true$ at an earlier step, with an error of at most $\frac{1}{1+\gamma}$. In other words, $<b^{\prime}_1, \cdots, b^{\prime}_n>$ $\Longleftrightarrow$ $<c^{\prime}_1, \cdots, c^{\prime}_n>$ in Algorithm~\ref{fptas} with an error of at most $\frac{1}{1+\gamma}$.

\noindent
We observe that $T^{\prime}_2[j] = T^{\prime}_1[j]$ for all $j \in [n]\setminus\{i\}$ . For $j=i$ we have $T^{\prime}_2[i] = b^{\prime}_i = \psi(b_i) = max(0,\psi(c_i - r_{im})) \geq \psi(c_i - r_{im})$. Algorithm~\ref{fptas} would calculate the value of position $i$ for vector $W^{\prime}_i$ as $W^{\prime}_i[i] = max(0, \psi(c^{\prime}_i-r_{im})) \geq \psi(c^{\prime}_i - r_{im})$.  
\noindent
We have $\psi(c^{\prime}_i - r_{im}) = \psi(\psi(c_i) - r_{im}) \geq \psi(\frac{c_i - r_{im}}{1+\varepsilon}) \geq \frac{1}{1+\gamma}\psi(c_i - r_{im})$, where $\gamma = 2 \varepsilon$.
\qed

\begin{lemma}\label{runtime}
Algorithm~\ref{fptas} has a runtime of $O(Bn\displaystyle\prod\limits_{i=1}^n log_{1+\varepsilon}({b_i}^{max}+1))$.  
\end{lemma}

\begin{lemma}\label{ratio}
The value of the solution $(S_{fptas})$ returned by Algorithm~\ref{fptas} differs from that $(S_{opt})$ returned by Algorithm~\ref{algo-dyn-prog} at most by a factor of $\frac{1}{1+\varepsilon B}$.
\end{lemma}
\proof At any step $i$, the value of any position of any vector of Algorithm~\ref{fptas} differs from the corresponding position of the corresponding vector for Algorithm~\ref{algo-dyn-prog} by a factor of $\frac{1}{(1+\varepsilon)^{i}} \approx \frac{1}{1+\varepsilon i}$ due to the rounding. We perform this rounding $B$ times. Hence the values of the vectors after the full execution would differ by a factor of at most $\frac{1}{1+\varepsilon B}$.  
\qed
\noindent
Lemma~\ref{runtime} and~\ref{ratio} lead to the following theorem.

\begin{theorem}
Algorithm~\ref{fptas} is an FPTAS for  \textsc{LMVT Problem}. 
\end{theorem}

\end{document}